\documentclass[11pt,tightenlines,aps,amsmath,amssymb,prb,superscriptaddress,reprint]{revtex4-1}

\usepackage{graphicx}
\usepackage{dcolumn}
\usepackage{bm}
\usepackage{times}
\usepackage{amsfonts}
\usepackage{graphicx}
\usepackage[table]{xcolor}
\usepackage[english]{babel}
\usepackage{color}

\begin{document}

\title[Low-Energy Truly Random Number Generation with Superparamagnetic Tunnel Junctions for Unconventional Computing]{Low-Energy Truly Random Number Generation with Superparamagnetic Tunnel Junctions for Unconventional Computing}

\author{D. Vodenicarevic}
\author{N. Locatelli}
\affiliation{Centre for Nanoscience and Nanotechnology, CNRS, Univ. Paris-Sud, Universit\'e Paris-Saclay, C2N Orsay, 91405 Orsay, France}
\author{A. Mizrahi}
\affiliation{Centre for Nanoscience and Nanotechnology, CNRS, Univ. Paris-Sud, Universit\'e Paris-Saclay, C2N Orsay, 91405 Orsay, France}
\affiliation{Unit\'e Mixte de Physique CNRS, Thales, Univ. Paris-Sud, Universit\'e Paris-Saclay, 91767 Palaiseau, France}
\author{J. S. Friedman}
\affiliation{University of Texas at Dallas, 800 West Campbell Road Richardson, TX 75080, USA}
\author{A. F. Vincent}
\affiliation{Centre for Nanoscience and Nanotechnology, CNRS, Univ. Paris-Sud, Universit\'e Paris-Saclay, C2N Orsay, 91405 Orsay, France}
\author{M. Romera}
\affiliation{Unit\'e Mixte de Physique CNRS, Thales, Univ. Paris-Sud, Universit\'e Paris-Saclay, 91767 Palaiseau, France}
\author{A. Fukushima}
\author{K. Yakushiji}
\author{H. Kubota}
\author{S. Yuasa}
\affiliation{AIST Tsukuba, 1-1-1 Higashi, Tsukuba, Ibaraki 305-8561, Japan}
\author{S. Tiwari}
\affiliation{School of ECE, Cornell University, Ithaca, New York 14850, USA}
\author{J. Grollier}
\affiliation{Unit\'e Mixte de Physique CNRS, Thales, Univ. Paris-Sud, Universit\'e Paris-Saclay, 91767 Palaiseau, France}
\author{D. Querlioz}
\email{damien.querlioz@u-psud.fr}
\affiliation{Centre for Nanoscience and Nanotechnology, CNRS, Univ. Paris-Sud, Universit\'e Paris-Saclay, C2N Orsay, 91405 Orsay, France}

\date{\today}

\begin{abstract}
Low-energy random number generation is critical for many emerging computing schemes proposed to complement or replace von Neumann architectures. 
However, current random number generators are always associated with an energy cost that is prohibitive for these computing schemes.
In this paper, we introduce random number bit generation based on specific nanodevices: superparamagnetic tunnel junctions. We experimentally demonstrate high quality random bit generation that represents orders-of-magnitude improvements in energy efficiency compared to current solutions. We show that the random generation speed improves with nanodevice scaling, and investigate the impact of temperature, magnetic field and crosstalk. 
Finally, we show how alternative computing schemes can be implemented using superparamagentic tunnel junctions as random number generators. 
These results open the way for fabricating efficient hardware computing devices leveraging stochasticity,
and highlight a novel use for emerging nanodevices.
\end{abstract}

\maketitle

\section{Introduction}

With conventional transistor technology reaching its scalability limits~\cite{courtland_transistors_2016}, significant effort is involved in the investigation of alternative computing schemes for microelectronics.
Many of these emerging ideas, such as stochastic computing~\cite{alaghi_survey_2013,friedman_bayesian_2016,morro2015ultra,hamilton2014stochastic,winstead2005stochastic} and some brain-inspired (or neuromorphic) schemes~\cite{merolla_million_2014,maass2014noise,suri2013bio}, 
require a large quantity of random numbers.
However, the  circuit area and the energy required to generate these random numbers are major limitations of such computing schemes. 
For example, in the popular neuromorphic TrueNorth system~\cite{merolla_million_2014}, one third of the neuron area is dedicated to perform random number generation.
Indeed, one million random bits are required, at each integration step of the system.
More concerning, in stochastic computing architectures, random number generation is typically the dominant source of energy consumption, as the logic performed using the random bits is generally quite simple and efficient by principle.
Many practical stochastic computing schemes therefore try to limit the reliance on expensive independent random bits using various techniques, including the sharing or reuse of random bits \cite{brown_stochastic_2001,morro_ultra-fast_2015,tehrani_stochastic_2006}. 
However, such tricks limit the capabilities of stochastic computing to small tasks, as they introduce correlations between signals.

Most of the aforementioned unconventional computing circuits use pseudo-random number generators. 
But these either lead to low quality random numbers or are highly energy and area-consuming.
A preferable solution would be to rely on ``true'' random number generators that generate random bits based on physical phenomena that are intrinsically random.
However, this is also difficult to realize with minimal energy consumption. 
This difficulty is due to the fact that
most true random number generators function by triggering events whose outcome is intrinsically random.
Triggering these events comes with a non-negligible energy cost.
The most energy-efficient example uses a bistable CMOS circuit forced into in a meta-stable state which then randomly falls into one of the two stable states, generating one random bit~\cite{mathew_2.4_2012}. 
It consumes $3\mathrm{pJ}/\mathrm{bit}$ and a circuit area of $4000\mathrm{\mu m}^2$.

In order to reduce this large area footprint, recent proposals suggest to leverage the inherent stochastic programming properties that arise in many of the bi-stable nano-devices developed for memory applications~\cite{rajendran_nano_2015}. This approach was investigated with oxide-based resistive memory devices~\cite{huang_contact-resistive_2012, balatti2015true, wang_novel_2015, hu_leveraging_2016}, phase-change memory devices~\cite{piccinini_self-heating_2017, fong_generating_2014}, magnetic memory devices~\cite{fukushima_spin_2014, choi_magnetic_2014, oosawa_design_2015}, as well as with straintronic memory devices~\cite{barangi_straintronics-based_2016}.
However, these approaches are based on repeated, energy-intensive programming operations, and still require high energy for random bit generation. For instance, it requires dozens of $\mathrm{pJ}/\mathrm{bit}$ to induce a stochastic switch of magnetization in magnetic tunnel junctions with two stable states, as proposed in the "Spin-Dice" concept, due to the high energy barrier between the magnetic states.
Optimized schemes have been proposed~\cite{kim_spin-orbit-torque-based_2015, sengupta_true_2016, lee_design_2017}, 
predicting further reduction of the energy cost per bit, but are still bounded by the need of a costly perturb operation.
While proposing high quality random number with high throughput, such strategies are no fit for emerging neuro-inspired computing applications like stochastic computing architectures.

A more natural approach would be to extract random numbers directly from thermal noise, as it provides randomness at no energy cost. 
Unfortunately,  this approach requires large circuits to amplify thermal noise into a large signal of random bits, and has never been shown to be more energy efficient than the first approach until now. 
The lowest energy solution today is to use jitter as a way to efficiently amplify the noise present in CMOS ring oscillators. 
The most energy efficient implementation \cite{yang_16.3_2014} requires $23pJ/bit$ and $375\mu m^2$.

In the present work, we propose the use of a nanomagnetic device that intrinsically amplifies thermal noise without external energy supply: superparamagnetic tunnel junctions. These bi-stable magnetic tunnel junctions are reminiscent of the ones used for Magnetic Random Access Memories (MRAMs) \cite{apalkov2013}. However, contrarily to MRAMs cells, the energy barrier between the two magnetic states is very low, and thermal fluctuations induce repeated and stochastic magnetization switching between the two states at room temperature.
Therefore, no write operations are required and a low-energy readout of the device state naturally produces random bits. 
We show that these devices permit the generation of high quality random numbers at $20\mathrm{fJ}/\mathrm{bit}$ using less than $2\mathrm{\mu m}^2$, which is orders of magnitude more efficient in terms of energy and area than current solutions.

We first show experimentally that superparamagnetic tunnel junctions allow the generation of high-quality random bits with minimal readout circuitry and that their behavior can be predicted by existing physical models.
We then use the model to investigate the influence of device scaling and environmental factors on random bit quality and speed. 
Circuit simulation enables an estimation of the energy efficiency of random bit generation.
Finally, we demonstrate the potential of these devices for unconventional computing through the example task of email messages classification using random bits extracted from the experimental data, and show that they are particularly adapted to computing schemes trading off speed for ultra low energy consumption.

\section{Exploiting the Stochastic Behavior of Superparamagnetic Tunnel Junctions}

\begin{figure}
\centering
\includegraphics[width=1\linewidth]{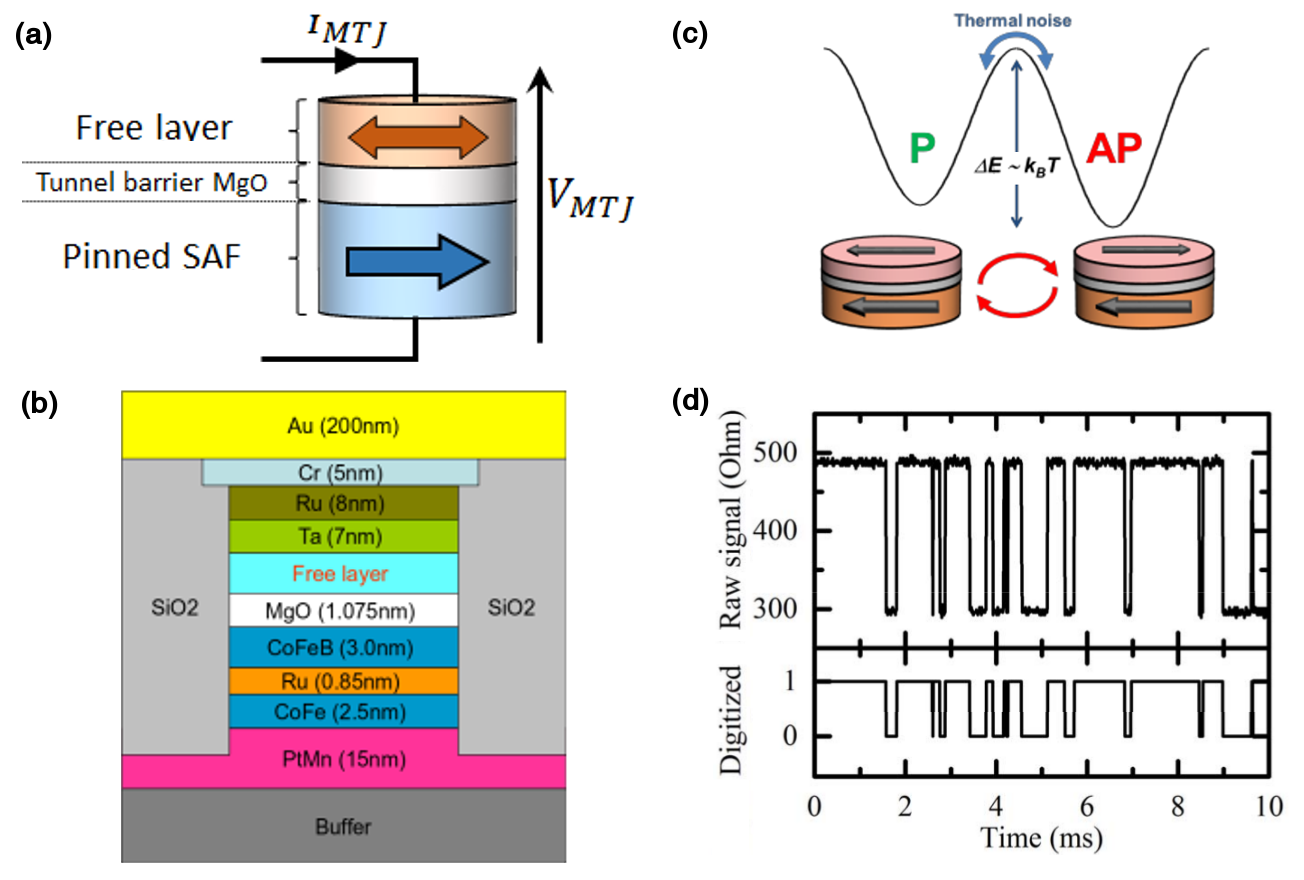}
\caption{\textbf{Structure and behavior of superparamagnetic tunnel junctions.} \textbf{(a)} Basic structure of the measured superparamagnetic tunnel junctions and readout setup. \textbf{(b)} Detailed stack of the junctions. \textbf{(c)} Representation of the two stable magnetic states, and the associated energy barrier. \textbf{(d)} Experimental resistance trace and thresholding operation.}
\label{fig1}
\end{figure}

Superparamagnetic tunnel junctions are bistable spintronic nanodevices composed of a high stability pinned nanomagnet and a low-stability ``free'' nanomagnet, separated by a tunnel oxide layer (Fig.~\ref{fig1}(a)). Their structure is highly similar to the magnetic tunnel junctions used as the basic cells of MRAMs. The devices we measured were fabricated by sputtering, with a standard magnetic tunnel junction process, with the CMOS-compatible stack detailed in Fig.~\ref{fig1}(b). E-beam lithography patterning was then performed to produce $50\times 150nm^2$ elliptic pillars.

The free magnet has two stable states, parallel (P) and antiparallel (AP) relatively to the pinned layer (Fig.~\ref{fig1}(c)). Through the tunnel magneto-resistance effect \cite{sun_magnetoresistance_2008}, the electrical resistance of the junction in the AP state $R_{AP}$ is higher than the resistance in the P state $R_P$. This effect is traditionally measured through the $\mathrm{TMR}$ coefficient defined by $R_{AP}/R_P=1+\mathrm{TMR}$. 

The lateral dimensions of the device are chosen so that the effective energy barrier between the two stable states is not very high compared to $k_B T$.
Unlike the case of MRAMs, for which the magnetization direction of the free magnet is highly stable and can only be switched by proper external action, the magnetization direction of the superparamagnetic free magnet spontaneously switches between its two stable states, due to low stability relative to thermal fluctuations (Fig.~\ref{fig1}(c)) \cite{rippard_thermal_2011, mizrahi2016controlling}. Here, no bias or perturb scheme is required to provoke these random fluctuations, but only temperature.

Resistance versus time measurements were done on junctions by applying a small $10\mu A$ constant current through the junction. Such a small current amplitude was chosen to have negligible influence on the magnetic behavior of the device\cite{vincent_analytical_2015} and to maximize its lifetime while providing a clear signal. 
Fig.~\ref{fig1}(d) shows a sample from the time evolution of the electrical resistance of a junction measured at room temperature, as well as a binarized version, obtained by thresholding.
We see that the resistance follows two-state fluctuations analogous to a random telegraph signal. The mean frequency of fluctuations is strongly related to the shape and material properties of the junction\cite{sato_properties_2014}.

\begin{figure*}
\centering
\includegraphics[width=1\linewidth]{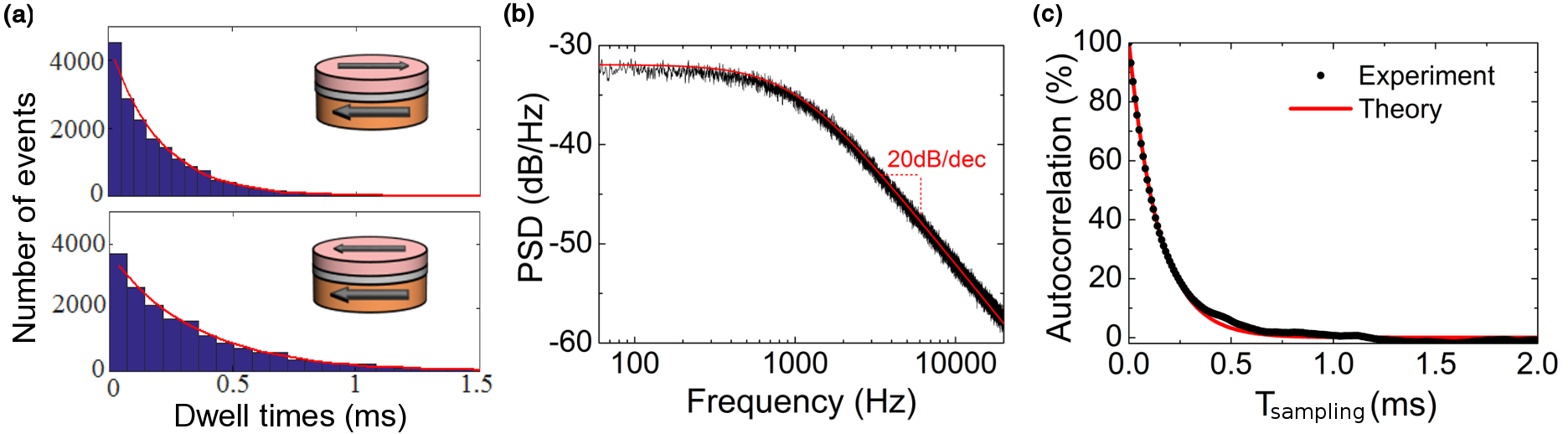}
\caption{\textbf{Statistics of the experimental superparamagnetic tunnel junction signal. }\textbf{(a)} Experimental histograms of the dwell times in Anti-parallel (AP, top Figure, high resistance) and Parallel (P, bottom Figure, low resistance) states, for a superparamagnetic magnetic tunnel junction measured during $10s$. \textbf{(b)} Experimental power spectrum density of the resistance signal. \textbf{(c)} Autocorrelation of the experimental resistance signal as a function of signal sampling period.}
\label{fig2}
\end{figure*}

Fig.~\ref{fig2}(a) shows the histograms of the dwell times in the `1' (AP) and `0' (P) states, obtained through measurement of a superparamagnetic tunnel junction over a 10 second period. 
We see that these histograms can be fitted by an exponential law, which is characteristic of a Poisson process. 
Fig.~\ref{fig2}(b) presents the power spectrum density of the same signal, superimposed with the expected power spectrum density of a random telegraph  signal based on a Poisson process. Excellent agreement between the measured results and the hypothesis of a Poisson process is seen.

Random bits can be extracted by sampling the voltage across the device at a constant frequency. 
The voltage was initially sampled at $100\mathrm{kHz}$, and bitstreams with slower sampling rates were obtained by subsampling the initial bitstream.
To evaluate the quality of the obtained random bits, the device was measured for over $2.5$ days, producing $21.2$ gigabits. No external magnetic field was applied during the measurement.

\section{Optimizing the quality of random bits}

The sampling frequency needs to be chosen carefully relative to the mean switching frequency of the junction, defined as $F_\mathrm{MTJ}=1/(\tau_\mathrm{1} + \tau_\mathrm{0})$, where $\tau_\mathrm{1}$ and $\tau_\mathrm{0}$ are the mean dwell times in state $1$ and $0$, respectively.
$F_\mathrm{MTJ}$ was measured to be $1.66\mathrm{kHz}$ ($\tau_\mathrm{1} + \tau_\mathrm{0} \approx 604\mu s$).
Fig.~\ref{fig2}(c) presents the correlation of consecutive bits extracted at different sampling rates.
This result is superimposed with the one theoretically expected from a Poisson process. 
At high sampling frequency, subsequent bits are naturally autocorrelated (at  $F_\mathrm{sampling}=100\mathrm{kHz}$, correlation reaches $92.8\%$), and can therefore not be used for applications. 
This correlation decreases exponentially with the sampling period, which can therefore be chosen based on the  correlation requirements on the random numbers.

As observed, in Fig.~\ref{fig2}(a),  the AP and P states possess an  asymmetric stability: the device spends more time on average in the P state than in the AP state, which corresponds to a mean state (mean of the binarized signal) of $60.5\%$. 
This asymmetry can be connected to the stray field induced by the pinned magnet layer structure, which is present in all magnetic tunnel junctions \cite{hayakawa_current-induced_2006}. 
This biasing field offsets the junction mean state from the ideal $50\%$ value required for most applications, and is subject to device-to-device variations.  

In order to eliminate this bias and any residual bit correlation, a ``whitening'' of the random bits is therefore required. 
To achieve this operation, we make use of a standard technique: combining several bitstreams into a single one  using XOR gates. 
It can be shown (mathematical derivation available in supplementary information S10) that the auto-correlation after XOR whitening is the product of the individual auto-correlations of the combined signals. It therefore decreases exponentially with the number of combined MTJ bitstreams, and is always lower than the auto-correlation of any of the combined signals. In the same way, the mean state of the whitened bitstream gets exponentially closer to $50\%$ with the number of XOR-combined bitstreams, and stays always closer to perfect balance than any of the bitstreams being combined.
As a reference, a more advanced but heavy stateful whitening technique ("Blum" \cite{blum_independent_1986}) was also applied to the raw measurements.

\begin{figure*}
\centering
\includegraphics[width=1\linewidth]{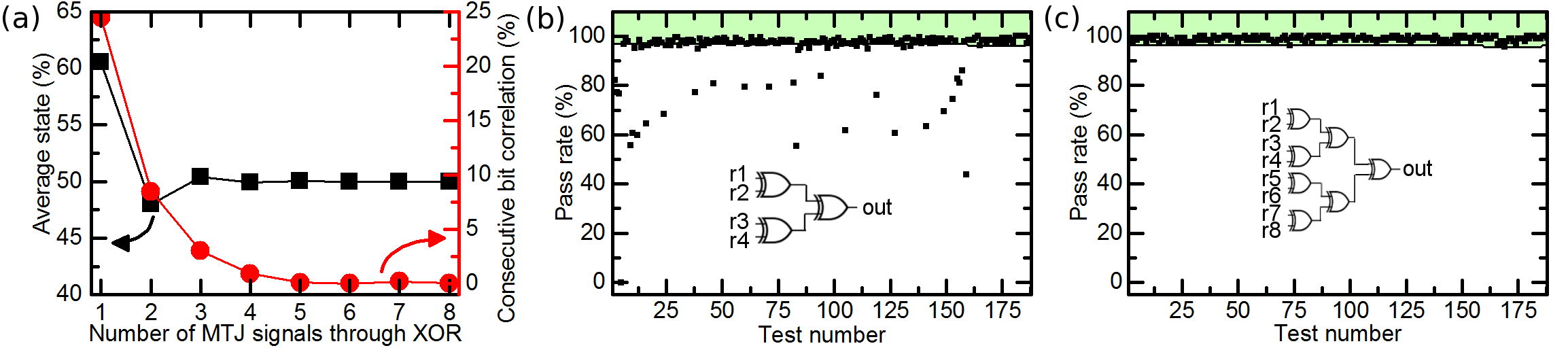}
\caption{\textbf{Whitened experimental random bitstream quality assessment. }\textbf{(a)} Mean state and consecutive bit autocorrelation as functions of the number of independent superparamagnetic tunnel junction signals combined by XOR. NIST STS randomness quality test results on experimental  data whitened by XOR4 \textbf{(b)} and XOR8 \textbf{(c)} at a $F_\mathrm{sampling}=5\mathrm{kHz}$ sampling frequency. When all test results are in the green area, the bitstream is consistent with cryptographic quality.}
\label{fig3}
\end{figure*}

As an illustration, we consider bits extracted at a frequency of $5\mathrm{kHz}$. The bitstream was then divided into chunks of equal length which were used as independent signals and XOR-combined bit per bit for the XOR whitening process.
We plot in Fig.~\ref{fig3}(a) the consecutive bit correlation and the mean state of the whitened bitstream as functions of the number of signals combined by XOR.
The correlation and the mean value bias decrease with the number of XOR-combined signals. With 4 bitstreams (XOR4), the resulting consecutive bit correlation drops under $1\%$ and the mean value reaches $49.9\%$. 
For 8 bitstreams (XOR8), the auto-correlation is below $0.06\%$ and the mean state reaches $50\%$ with a standard deviation of $0.5\%$. 
These results suggest that XOR whitening can correct correlation and mean value issues.

However, in order to fully evaluate the quality of a whitened bitstream, signal autocorrelation and mean state are not sufficient metrics. 
We therefore used the standardized National Institute of Standards and Technology Statistical Test Suite (NIST STS) \cite{soto_statistical_1999}, which evaluates the quality of the random bitstream against 188 tests. 
The NIST STS computes the statistics of bitstreams, such as mean value, auto-correlation, standard deviation, estimated entropy or pattern occurrence frequencies, and checks weather they are consistent with perfect randomness. It also looks for the presence of repeated structures, linear dependencies, and other behaviors unexpected in a perfectly random bitstream.

To perform the NIST STS tests, the bitstream to be tested, measured during 2.5 days, is divided into 1~Mbits sequences. Each chunk is then tested independently, and the pass rate (percentage of one million bits sequences passing the test) was computed for each of the 188 tests. Fig.~\ref{fig3}(b) and (c) show the results for XOR4 and XOR8 whitened bitstreams respectively.
For a bitstream to be consistent with cryptographic quality, the pass rates of all tests should lie in the green region \cite{soto_statistical_1999}, corresponding to the expected minimal pass rate provided by the NIST STS, dependent on the number of tested chunks.
We can see that bits extracted by XOR8 whitening pass this requirement (this was also the case with the reference Blum technique), while with XOR4 whitening only a fraction of the tests are consistent with cryptographic quality of the random bits
\footnote{The NIST tests also include a uniformity condition on the distribution of P-Values among tested sequences\cite{soto_statistical_1999}. This condition was passed for all tests for sequences processed by Blum and XOR8.}.

\begin{table}
\begin{center}
\begin{tabular}{|c|c|c|c|c|c|}
  \hline
   \cellcolor{gray!20}\textbf{$~F_\mathrm{sampling}~$} & \cellcolor{gray!20}\textbf{$F_\mathrm{sampling}/F_\mathrm{MTJ}$} & \cellcolor{gray!20}\textbf{~~Raw~~} & \cellcolor{gray!20}\textbf{XOR2} & \cellcolor{gray!20}\textbf{XOR4} & \cellcolor{gray!20}\textbf{XOR8} \\
  \hline
  100 kHz&    60.4	&\cellcolor{red!50}0	&\cellcolor{red!25}10.1	&\cellcolor{red!25}10.1	&\cellcolor{red!25}10.1 \\
  \hline
  20 kHz&    12.1	&\cellcolor{red!50}0.5	&\cellcolor{red!50}0.5	&\cellcolor{red!25}10.6	&\cellcolor{red!25}12.2 \\
  \hline
  9.1 kHz&    5.5		&\cellcolor{red!50}1.1	&\cellcolor{red!25}10.6	&\cellcolor{red!25}10.6	&\cellcolor{orange!25}88.3 \\
  \hline
  5.9 kHz &   3.6		&\cellcolor{red!50}1.1	&\cellcolor{red!50}1.1	&\cellcolor{red!25}16.5	&\cellcolor{green!50}100 \\
  \hline
  5 kHz&      3.0		&\cellcolor{red!50}1.1	&\cellcolor{red!50}1.1	&\cellcolor{orange!50}72.9	&\cellcolor{green!50}100 \\
  \hline
  1.9 kHz&    1.1		&\cellcolor{red!50}1.1	&\cellcolor{red!25}14.4	&\cellcolor{green!25}97.9	&\cellcolor{green!50}100 \\
  \hline
  0.9 kHz&    0.54		&\cellcolor{red!50}1.1	&\cellcolor{red!25}14.4	&\cellcolor{green!25}98.4	&\cellcolor{green!50}100 \\
  \hline
  0.7 kHz&    0.42		&\cellcolor{red!50}1.1	&\cellcolor{red!25}16.0	&\cellcolor{green!25}97.9	&\cellcolor{green!50}100 \\
  \hline
  0.5 kHz&    0.30		&\cellcolor{red!50}1.1	&\cellcolor{red!25}16.0	&\cellcolor{green!25}98.4	&\cellcolor{green!50}100 \\
  \hline
\end{tabular}
\end{center}
\caption{\textbf{NIST Statistical Test Suite results for the whitened experimental random bitstream. }Percentage of NIST STS tests satisfying cryptographic quality requirements for different numbers of combined bitstreams, and different sampling frequencies.}
\label{tab1}
\end{table}

Table~\ref{tab1} presents more comprehensive results: the proportion of tests in the green area  is given  for XOR-whitened bitstreams at different sampling frequencies and numbers of XOR-combined bitstreams. 
The results confirm that the quality of the whitened bitstream increases for lower sampling frequencies (less correlation) and higher numbers of XOR-combined bitstreams (less correlation and lower bias). Higher numbers of XOR-combined bitstreams therefore allow further increasing the sampling rate while still passing all the NIST statistical tests, at the expense of more circuit area and energy consumption.
XOR8 at $F_\mathrm{sampling}/F_\mathrm{MTJ}=3.0$ appears to be an optimal choice, with 100\% of the tests consistent with cryptographic quality and the highest sampling frequency. 
A more comprehensive analysis of the impact of the number of XORed bitstreams is presented in supplementary information S1.

Consistent results (presented in supplementary information S2) were observed on a second sample, measured during 1.5 days, producing $8.96$ gigabits.

\section{Scaling Capabilities of the Random Number Generators in Terms of Speed and Energy Consumption}

\begin{figure*}
\centering
\includegraphics[width=1\linewidth]{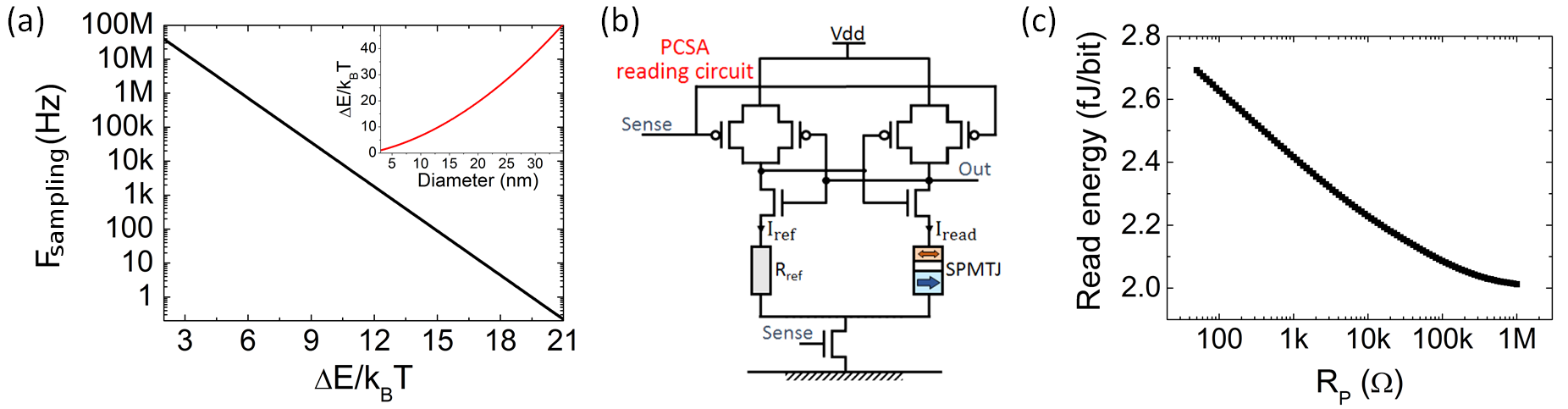}
\caption{\textbf{Sampling rate and readout circuitry. }\textbf{(a)} Effect of scaling the energy barrier on the ideal sampling frequency, based on the device model. The inset shows the energy barrier as a function of the junction diameter for PMA-MTJs. \textbf{(b)} Precharge sense amplifier (PCSA) circuit for reading the state of a superparamagnetic tunnel junction. \textbf{(c)} PCSA reading energy as a function of the superparamagnetic tunnel junction P state resistance $R_P$.}
\label{fig4}
\end{figure*}

Further studying the potential of superparamagnetic tunnel junctions for random number generation requires a realistic model of the device. 
In the literature, at low electric current, magnetic tunnel junctions switching is usually described by an Arrhenius-N\'eel two-states analysis, modeling a thermally activated magnetic switching \cite{mizrahi_magnetic_2015}.
The mean switching rates in each state are then described by: 
\begin{equation}
  \begin{cases}
        r_{0 \rightarrow 1} = 1/\tau_{0} = f_0 \exp\left(-\frac{\Delta E_{0 \rightarrow 1}}{k_B T}\right) \\
        r_{1 \rightarrow 0} = 1/\tau_{1} = f_0 \exp\left(-\frac{\Delta E_{1 \rightarrow 0}}{k_B T}\right)
  \end{cases}
\end{equation}
where $f_0=1\mathrm{GHz}$ is the magnetic attempt frequency, $\Delta E_{0 \rightarrow 1}$ and $\Delta E_{1 \rightarrow 0}$ are the energy barriers associated with each transition (see Fig.~\ref{fig1}(c)).
Our experimental results suggesting that superparamagnetic tunnel junction switching is a Poisson process are consistent with this model.

The superparamagnetic tunnel junctions that we characterized experimentally in this study are  slow devices. 
They can be used to generate random bits at $\mathrm{kHz}$ frequencies, sufficient for real-time brain-inspired systems like \cite{merolla_million_2014}, but not for high performance applications.
In  our $50\times 150\mathrm{nm}$ superparamagnetic tunnel junctions, we identified that the switching occurs through nucleation and propagation of a magnetic domain, probably seeded by fluctuations in a subset of grains within it\cite{rippard_thermal_2011} (supplementary information S3). 
By contrast, recent experiments on perpendicular magnetic anisotropy (PMA) magnetic tunnel junctions 
have shown that aggressively scaled devices (diameters smaller than $35\mathrm{nm}$) switch at the scale of the whole volume \cite{sato_properties_2014}.
Therefore, in the context of random number generators, extreme scaling of the nanodevices appears as providential, as smaller volumes and areas are directly linked to a lower magnetization stability of the free magnet \cite{sato_comprehensive_2013},  increasing random bit generation speed exponentially.
This is in sharp contrast with MRAMs, where conservation of stability with extreme scaling is an important challenge \cite{maffitt2006design}.

From the study described in previous part, we observe that a $25\%$ correlation between consecutive bits can be efficiently whitened out by XOR8 and allow generated random numbers to pass all the NIST STS tests. 
This consideration, together with the model, allows us to evaluate quantitatively the speed of scaled random bit generators based on superparamagnetic tunnel junctions, by evaluating the maximum sampling frequency to keep the correlation $\rho_{X,X+1}^{c} \lesssim 25\%$ (details in supplementary information S4):
\begin{equation}
F_\mathrm{sampling}^\mathrm{max} \approx 3 F_\mathrm{MTJ} = \frac{3}{2}f_0\exp\left(-\frac{\Delta E}{k_B T}\right)
\end{equation}
\noindent where $\Delta E$ is the energy barrier separating the two states.
$\Delta E = K_\mathrm{eff}(D) \pi \frac{D^2}{4} t$ is derived as a function of the device diameter D, where $t=1.6$nm is the free magnet thickness and the effective anisotropy $K_{eff}(D)$ is derived considering interfacial anisotropy and bulk anisotropies, using experimental values from \cite{sato_properties_2014}.
Fig.~\ref{fig4}(a), based on this derivation, shows that random bits could be generated at up to tens of $\mathrm{MHz}$ for energy barriers below $5k_BT$, corresponding to a diameter of $8\mathrm{nm}$.

In addition, in a final system, specialized transistor-based electronics needs to be associated to the superparamagnetic tunnel junctions to read their states without interfering with the random bit generation quality. 
Here, we considered a precharge sense amplifier circuit (PCSA, Fig.~\ref{fig4}(b)), a CMOS circuit originally proposed as an MRAM read circuit \cite{zhao_high_2009}. 
We simulated this circuit using standard integrated circuit design software (Cadence tools) and  the transistor models of  a $28$nm commercial technology. The superparamagnetic tunnel junctions were modeled using a compact (VerilogA-based) model implementing the Arrhenius-N\'eel model. 
The results of circuit simulation (Fig.~\ref{fig4}(c)) show that the read energy is relatively independent from superparamagnetic tunnel junction resistance, and very low ($\approx 2\mathrm{fJ/bit}$).
We also evaluated the read disturb effect of the PCSA.
Reading the state of a junction can potentially affect random bit generation through the spin torque effect.
Based on the spin torque model of Ref.~\onlinecite{mizrahi_magnetic_2015},
its impact on the mean state is around $10^{-6}$ for junctions such as the one we characterized experimentally. 
It would stay below $0.1\%$ for ultrascaled junctions functioning at high frequencies, as shown in supplementary material S11.
This small effect would therefore be corrected by whitening. 

Evaluating the energy consumption of random bit generation requires taking into account the whitening process.
As XOR whitening combines multiple junction states per generated bit, it requires multiple read operations per generated bit. XOR8 reads 8 junctions to generate a bit, and requires 20fJ/bit in average (including the XOR gate operation).  
In terms of area, in a 28~nm technology, the layout of a full XOR8 random bit generator takes less than $2\mu m^2$. 
XOR4 whitening would require 9.8fJ/bit and a $1\mathrm{\mu m}^2$ area.

These results show the potential of superparamagnetic tunnel junctions for state of the art low-energy random number generation.

\section{Sensitivity of the Random Number Generators to Perturbations}

\begin{figure*}
\centering
\includegraphics[width=1\linewidth]{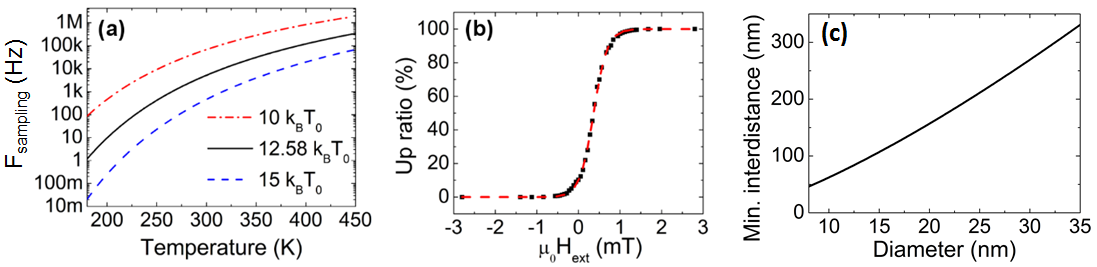}
\caption{\textbf{External perturbations and crosstalk effects. }\textbf{(a)} Theoretical curve of the maximum sampling frequency for high quality random bit generation, as a function of temperature, for different junction stabilities (the black curve corresponds to the junction characterized in Figs.~\ref{fig2} and ~\ref{fig3}). \textbf{(b)} Black symbols: experimental mean state of the junction (up ratio) as a function of the applied magnetic field (red dotted line: theory). \textbf{(c)} Theoretical minimal distance between superparamagnetic tunnel junctions allowed to prevent crosstalk, as a function of the superparamagnetic tunnel junction diameter.}
\label{fig5}
\end{figure*}

Although superparamagnetic tunnel junctions allow random number generation with minimal energy, their sensitivity to external perturbations must be carefully evaluated.

First, as the stochastic switching of superparamagnetic tunnel junctions is thermally activated, temperature directly affects their switching rates. 
Fig.~\ref{fig5}(a), based on the model introduced in the previous section, shows the temperature dependence of the maximum sampling frequency for several values of the effective barrier. 
Higher temperatures produce better random numbers: as temperature increases, the superparamagnetic tunnel junction switching rates increase accordingly, thus allowing faster sampling frequencies. 
Devices should therefore be sized based on their lowest operation temperature.

Superparamagnetic tunnel junctions are also sensitive to magnetic fields.
Fig.~\ref{fig5}(b) shows the experimental mean state of a superparamagnetic junction as the function of external magnetic field.
Fields of  a few Oe shift the mean state to a level that cannot be corrected by XOR8 whitening. Magnetic shielding  is therefore necessary for applications.
Such technology  (based on mu metals) has already been developed for MRAM.

Finally, a challenge regarding scalability and integration is that closely packed superparamagnetic tunnel junctions can interact by dipolar interaction, which could lead to correlations in random numbers.
In the case of perpendicularly magnetized superparamagnetic tunnel junctions, using the previously introduced model, we evaluated that the critical center-to-center distance between two superparamagnetic tunnel junctions guaranteeing negligible crosstalk \cite{neiman_synchronization_1999}, corresponding to less than $\rho_{c} = 0.1\%$ cross-correlation, is given by (details in supplementary information S5):
\begin{equation}
d_{c} = \left( \frac{\mu_{0}(M_{S}V)^{2}}{4\pi k_{B} T \tanh^{-1}(\rho_{c})} \right)^{1/3} 
\end{equation}

\noindent Fig.~\ref{fig5}(c) shows the evolution of this critical distance at room temperature as the diameter of the junctions is scaled down. It falls below $100\mathrm{nm}$ for ultimately scaled $10\mathrm{nm}$ diameter devices,  which constitutes a layout design rule, and which would be naturally respected if the junctions are associated with PCSA circuits.

\section{Using Superparamagnetic Tunnel Junctions for Unconventional Computing}

\begin{figure*}
\centering
\includegraphics[width=1\linewidth]{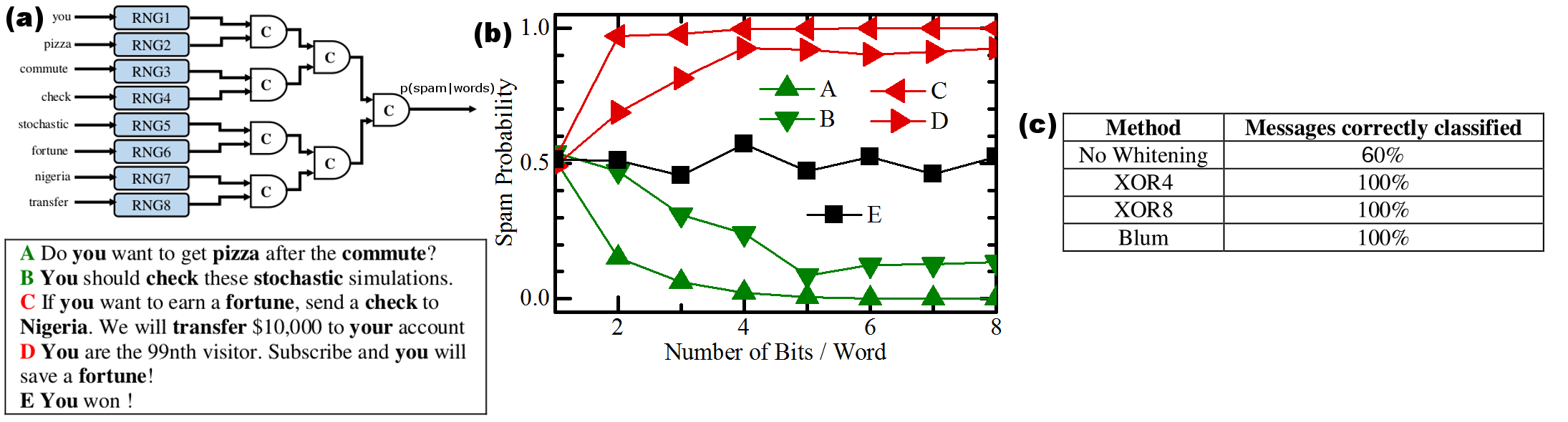}
\caption{\textbf{Email classification with stochastic computing using whitened experimental random bitstreams. }\textbf{(a)} Stochastic email classification circuit, and email messages to classify. One ``RNG'' block includes several random bit generators to provide bits with controllable probability. \textbf{(b)} Resulting spam probability as a function of the number of random bits per word using XOR4-whitened experimental $5\mathrm{kHz}$ data over 2000 iterations. \textbf{(c)} Spam classification success rates for different whitening techniques for $5\mathrm{kHz}$ sampling, using 8bits/word and 2000 iterations.}
\label{fig6}
\end{figure*}

To illustrate the potential of superparamagnetic tunnel junctions for unconventional computing, we use the experimental whitened random bitstreams as inputs for a modern stochastic circuit (Fig.~\ref{fig6}(a) and Ref.~\onlinecite{friedman_bayesian_2016}) that performs Bayesian inference as a non-Turing machine. 
As a pedagogical task, we use this circuit to classify email messages as spam or not spam (sample messages are presented in Fig.~\ref{fig6}(a)), as recently introduced in Ref.~\onlinecite{friedman_bayesian_2016}. 

The approach uses a dictionary of known words with their associated occurrence rates in spam and non-spam messages. 
Each word of the dictionary has an associated probabilistic binary generator whose probability of drawing a 1 is set to different values depending on the presence (or absence) of the word in the presented sentence. 
As our random bit generators  provide bitstreams with mean values of 0.5, multiple random bit generators   are needed to create a probabilistic binary generator (see ``RNG'' block in Fig.~ \ref{fig6}(a), detailed in supplementary information Fig. S6). 
The outputs of these generators are then combined using C-Elements to perform an approximate Bayesian inference \cite{friedman_bayesian_2016}. The time average of the output gives the probability of the presented message being spam. 

Fig.~\ref{fig6}(b) gives the spam probability inferred using XOR4-whitened bitstreams and shows that the more random bit generators are used per word, the more precisely the probabilistic binary generator can be tuned and the better the prediction is. Also, the longer the output averaging time, the more accurate the answer of the system is. A trade-off to keep a low energy consumption is found for 8 random bit generators / word and averaging over 2000 samples (supplementary information S7 and S8).

Because of its reliance on multiple stages of binary bitstream combination, and fine generator probability tuning, this circuit is sensitive to the quality of the underlying random number generator. We tested the circuit using raw 5kHz-sampled experimental bitstreams, as well as its XOR4 and XOR8 whitened versions. When the bits are not whitened, the circuit does not perform satisfactorily (Fig.~\ref{fig6}(c) and supplementary information S7). Using bits whitened with XOR8, the circuit performs as well as the reference Blum whitener, successfully classifying all messages. Furthermore, XOR4, which does not pass all NIST STS tests, also provides perfect classification while requiring less energy.

These results highlight the potential of the approach for low-energy applications. Using the results of the previous section,  circuit simulation with 8 random bit generators / word and 2,000 clock cycles shows that a message can be classified using only nJ energy (the exact value depends on the number of words in the dictionary, see supplementary information S9). 
This simple study shows that superparamagnetic tunnel junctions can be used for efficient random number generation for low-power probabilistic computing.

\section{Conclusion}

In this work, we have shown that the natural dynamics of superparamagnetic tunnel junctions produces random telegraph signals that can be read and turned into high quality  random bitstreams with minimal energy and circuit overhead, while staying fully compatible with standard CMOS fabrication processes.

The whitening process turning these measurements into usable random bitstreams implies energy and area overhead. But while reference Blum whitening would add important CMOS overhead, XOR adds very little. XOR8 and Blum both provide high random bit quality  consistent with cryptographic requirements, but XOR8 fits better to low energy applications, as it typically requires only 20fJ/bit and $2\mu m^2$, orders of magnitudes less than current state of the art.
This efficiency comes at the cost of speed. Scaled superparamagnetic tunnel junctions could generate random bits at speeds of dozens of MHz, which is  slower than higher energy random bit generators, but sufficient for many unconventional computing schemes in very low power consumption contexts such as the Internet of Things.
This efficiency also comes at the cost of  a certain sensitivity of random bit generation to the environment, making it prone to attacks.
Random bit generation based on superparamagnetic tunnel junctions  is therefore much better suited for unconventional computing than for cryptographic applications.

The evaluation of the probabilistic email classifier circuit also suggests that in many alternative computing schemes, lower-quality whitening can be used successfully to achieve extreme energy efficiency without degrading performance. At design time, a balance between random number quality, generation speed, and energy consumption can be freely chosen to suit the target application. This is especially important in the context of modern Bayesian inference systems \cite{coninx2016bayesian,faix2016design}, but also for embedded circuits and Internet of Things applications that are designed to work at low frequencies and low energies.

This study shows, through the example of superparamagnetic tunnel junctions acting as natural noise amplifiers, that emerging nanodevices could be used as highly efficient sources of true randomness for a wide range of applications.

\begin{acknowledgments}
This work is supported by the European Research Council Starting Grant NANOINFER (reference: 715872), by the BAMBI EU collaborative FET Project grant (FP7-ICT-2013-C, project number 618024), by a public grant overseen by the French National Research Agency (ANR) as part of the Investissements d'Avenir program (Labex NanoSaclay, reference: ANR-10-LABX-0035), by the ANR grant CogniSpin (reference: ANR-13-JS03-0004), and by the French Minist\`ere de l'\'ecologie, du d\'eveloppement durable et de l'\'energie.\\
The authors thank J.~Droulez and P.~Bessi\`ere for fruitful discussion.
\end{acknowledgments}

\section*{Supplementary Materials}
\noindent
S1: Effect of whitening on random bitstream quality.\\
S2: NIST STS results on a second junction.\\
S3: Discussion on superparamagnetic junction switching process.\\
S4: Sampling frequency and correlation between consecutive samples.\\
S5: Crosstalk through dipolar interaction.\\
S6: Spam detector: Email classifier architecture.\\
S7: Effect of whitening on spam detection.\\
S8: Effect of the number of bits per word on spam detection.\\
S9: Energy consumption of spam detection.\\
S10: Effects of XOR whitening on bitstream probability and auto-correlation.\\
S11: Read disturb effect.\\

%

\end{document}